\documentclass[12pt,draftclsnofoot, onecolumn]{IEEEtran}

\usepackage[cmex10]{amsmath}
\usepackage{amsfonts,amssymb,cite,graphicx,delarray,color}
\usepackage{multirow}
\usepackage{url}
\IEEEoverridecommandlockouts

\newtheorem{remark}{Remark}
\newtheorem{lemma}{Lemma}
\newtheorem{proposition}{Proposition}

\begin{document}

\title{User-Centric Intercell Interference Nulling for Downlink Small Cell Networks}
\date{ }

\author{
    \IEEEauthorblockN{Chang~Li,~\IEEEmembership{Student Member,~IEEE}, Jun~Zhang,~\IEEEmembership{Member,~IEEE}, Martin~Haenggi,~\IEEEmembership{Fellow,~IEEE}, and Khaled~B.~Letaief,~\IEEEmembership{Fellow,~IEEE}}
    \thanks{C. Li, J. Zhang and K. B. Letaief are with the Dept. of ECE at the Hong Kong University of Science and Technology, Hong Kong (email: \{changli, eejzhang, eekhaled\}@ust.hk).}
    \thanks{M. Haenggi is with the Dept. of Electrical Engineering at the University of Notre Dame, IN, USA (email: mhaenggi@nd.edu).}
}

\begin{titlepage}
\maketitle
\thispagestyle{empty}
\vspace{-1.5cm}
\begin{abstract}
    Small cell networks are regarded as a promising candidate  to meet the exponential growth of mobile data traffic in cellular networks. With a dense deployment of access points, spatial reuse will be improved, and uniform coverage can be provided. However, such performance gains cannot be achieved without effective intercell interference management. In this paper, a novel interference coordination strategy, called \emph{user-centric intercell interference nulling}, is proposed for small cell networks. A main merit of the proposed strategy is its ability to effectively identify and mitigate the dominant interference for each user. Different from existing works, each user selects the coordinating base stations (BSs) based on the relative distance between the home BS and the interfering BSs, called the \emph{interference nulling (IN) range}, and thus interference nulling adapts to each user's own interference situation.  By adopting a random spatial network model, we derive an approximate expression of the successful transmission probability to the typical user, which is then used to determine the optimal IN range. Simulation results shall confirm the tightness of the approximation, and demonstrate significant performance gains (about 35\%-40\%) of the proposed coordination strategy, compared with the non-coordination case. Moreover, it is shown that the proposed strategy outperforms other interference nulling methods. Finally, the effect of imperfect channel state information (CSI) is investigated, where CSI is assumed to be obtained via limited feedback. It is shown that the proposed coordination strategy still provides significant performance gains even with a moderate number of feedback bits.
\end{abstract}

\vspace{-0.5cm}
\begin{keywords}
\vspace{-0.5cm} Poisson point process, stochastic geometry, intercell interference nulling, small cell networks, limited feedback.
\end{keywords}

\end{titlepage}

\baselineskip=24pt
\newpage

\section{Introduction} \label{Sec:Introduction}

\subsection{Motivation}

In the past few years we have witnessed an exponential growth of mobile data traffic, and this trend will continue \cite{Cisco14}. Significant efforts have been spent on increasing capacity of wireless networks to accommodate the mobile data tsunami. However, we are already approaching the Shannon limit of point-to-point links, and there is little extra radio spectrum to exploit. Recently, small cell networks have been proposed as a promising approach to address these challenges and further boost the network capacity. By deploying more access points, spatial reuse can be improved and more uniform coverage can be provided \cite{Hwang13}.

As the network gets denser, new design challenges arise, among which intercell interference management is a critical one. Without effective interference management, the performance of mobile users will be severely degraded by intercell interference from nearby base stations (BSs). For example, it was shown in \cite{Andrews11} that the outage probability of the typical user in a multi-cell network with Poisson distributed single-antenna BSs is higher than 40\% if the signal-to-interference-plus-noise ratio (SINR) threshold is 0 dB, even without additive thermal noise. The performance can be improved by deploying multi-antenna BSs. It was shown in \cite{Li14} that when each BS is equipped with 4 antennas, the outage probability with the SINR threshold as 0 dB can be reduced to below 10\% with maximum ratio transmission (MRT). However, without interference management, the performance will not be satisfactory when the SINR threshold increases, i.e., as the data rate requirement increases. Also in \cite{Li14}, it was shown that the outage probability with single-user beamforming (MRT) is about 40\% for an SINR threshold of 10 dB, even if each BS is equipped with 8 antennas. Therefore, to provide satisfactory user performance in dense small cell networks, effective interference management should be developed.

Recently, multi-cell cooperation has been proposed as an efficient way to mitigate intercell interference \cite{Gesbert10,Zhang09,LopezPerez11,Foschini06,Shamai04,Huh12,Bjornson13}. There are different types of cooperation strategies by assigning different temporal/spectral/spatial dimensions to users among different cells. \emph{Intercell interference nulling}, as one particular type of multi-cell cooperation, has been shown to be a practical and viable approach for downlink interference suppression  \cite{Zhang10,Bhagavatula11}. With interference nulling, user data is transmitted only from one BS, while control information is exchanged between BSs and thus the coordinating multi-antenna BSs can suppress interference to users in neighboring cells with interference nulling. Compared with joint precoding among BSs \cite{Zhang09}, interference nulling does not require data sharing between BSs and thus has a lower signaling overhead, which is more suitable for dense networks. Although the effectiveness of interference nulling has been well studied in small networks \cite{Zhang10,Zhang11,Bhagavatula11}, its application in a densely deployed network requires a detailed investigation, since there are new features when looking from a network level, such as irregular BS positions. In this paper, we will endeavor to develop an efficient yet low-complexity interference nulling strategy tailored for small cell networks and investigate its performance gain.

\subsection{Prior Works}

Most previous works on interference coordination either use the Wyner model \cite{Simeone09,Shamai01,Gesbert10}, or adopt the grid model \cite{Zhang10,Zhang11,Bhagavatula11,Foschini06,Papadogiannis08} with a finite number of cells. The Wyner model is oversimplified and does not capture the essential characteristics of real and practical networks \cite{Xu11}. For the grid model, the analysis becomes intractable as the network size grows, and thus simulation becomes a common approach to seek insights for the system design. Moreover, none of the above network models captures the irregular network structure in small cell networks. Recently, a random cellular network model was proposed in \cite{Andrews11}, where BSs are modeled as a spatial Poisson point process (PPP). This model captures the irregularity of the BSs and is about as accurate as the grid model while being much more tractable \cite{Guo13}.

Although there have been numerous studies using the PPP model to analyze cellular networks, e.g., \cite{Dhillon12,Cheung12,Singh13,Li13_GC,Elsawy13,Wang14}, most of them did not consider any interference coordination. This is mainly due to the difficulty of the performance analysis with cooperation among different BSs. There have been prior studies on interference management in cellular networks \cite{Akoum13,Huang13,Tanbourgi14,Nigam14,Zhang13,Xia13}. In \cite{Akoum13,Huang13}, all the BSs in the network are grouped into disjoint clusters, and each BS will avoid intercell interference to users in other cells within the same cluster with interference nulling beamforming. Joint transmission was investigated in \cite{Tanbourgi14,Nigam14}, where each user is served by several nearby single-antenna BSs under the assumption that the user data is shared between these BSs with high-capacity backhaul links. In \cite{Zhang13}, while intercell interference was avoided by serving users in different cells with orthogonal channels, intra-cell diversity was applied to further improve performance.

The disjoint BS clustering method \cite{Akoum13,Huang13} is designed from a transmitter's point of view and fails to consider each user's interference situation. Only users around the cluster center can benefit from such coordination, while the cluster edge users still suffer severe interference from neighboring clusters \cite{Zhang09}. To efficiently utilize the available radio resources, the coordinating BSs should be carefully selected to meet each user's demand. In \cite{Tanbourgi14,Nigam14,Zhang13,Xia13}, the set of coordinating BSs is determined from each user's point of view, which consists a fixed number of strong interferers. To make the analysis tractable, all of these studies assumed that each BS always has enough resources to handle all the coordinated users. However, such results may not be applicable to realistic networks, especially in small cell networks. On one hand, each BS has limited resources for interference suppression, e.g., with interference nulling, the number of interferers that can be handled is limited by the number of BS antennas. On the other hand, with irregularly placed BSs, different users will have different numbers of dominant interferers, and thus it is inefficient to enforce a fixed number of BSs for coordination. Therefore, a new criterion to effectively determine the coordinating BSs is needed to further improve the performance of interference coordination.

\subsection{Contributions}

In this paper, we will propose a novel \emph{user-centric intercell interference nulling} strategy for small cell networks. One main advantage of this strategy is that it can effectively determine the coordinating BSs for each user, which takes account of each user's interference situation and the limited resources at each BS. Specifically, each user will set an interference nulling (IN) range, based on its average received information signal power. The interfering BSs within the IN range are requested to do interference nulling for this user. The main design challenge is to specify the IN range: if it is too large, each BS may receive too many coordination requests, and thus it needs to spend most of its resources for interference nulling; if it is too small, the user will still suffer strong interference. In this paper, by adopting a random spatial network model, we analytically evaluate the successful transmission probability of the proposed strategy and determine the optimal IN range. Although the interference distribution becomes highly complicated with coordination, we develop a simple yet accurate approximate result.

Through numerical analysis, we compare the proposed interference nulling strategy with the non-coordination strategy, as well as other interference nulling methods, such as the random BS clustering method proposed in \cite{Akoum13}, and the user-centric coordination but with a fixed number of requests from each user \cite{Xia13}. We have the following findings: 1) The proposed strategy can greatly improve the successful transmission probability compared with the non-coordination case, and it outperforms other coordination methods, which indicates its effectiveness. 2) The proposed strategy provides a larger performance gain when the SINR threshold gets higher, which implies that it is capable to meet high data rate requirement. 3) To satisfy a given performance requirement for a certain user density, the proposed strategy needs much fewer BSs than the non-coordination strategy, which implies a significant reduction of the deployment cost.

Finally, we investigate the effect of imperfect CSI due to limited feedback. The approximate expression of the successful transmission probability is provided. We will then show that the performance of interference nulling depends critically on the number of feedback bits ($B$) for each channel vector. In particular, as $B$ increases, the performance gain from interference nulling becomes larger. If the feedback link has limited capacity, there exists a critical number of feedback bits below which it is better to use a non-coordination strategy.

\subsection{Paper Organization}

The rest of the paper is organized as follows. Section \ref{Sec:SystemModel} presents the system model and the proposed user-centric intercell interference nulling strategy. Section \ref{Sec:ps} derives the expression of the successful transmission probability. In Section \ref{Sec:LimitedFeedback}, we investigate the effect of limited feedback on the performance, while the numerical results and the comparison between different interference nulling methods are shown in Section \ref{Sec:NumResults}. Finally, Section \ref{Sec:Conclusions} concludes the paper.

\section{System Model} \label{Sec:SystemModel}

In this section, the random spatial model for small cell networks will be firstly presented, and then we will describe the user-centric intercell interference nulling strategy. Finally, we will introduce the performance metric used in this paper.

\subsection{The Network Model}

We consider a cellular network, where BSs and users are distributed in $\mathbb{R}^2$ according to two independent PPPs, denoted as $\Psi_{b}$ and $\Psi_{u}$, respectively. The density of BSs is denoted as $\lambda_b$ while the density of users is $\lambda_u$. We focus on the downlink transmission and assume that the BSs use the same transmit power $P_t$. Each user is served by the nearest BS, which implies that the cell of each BS corresponds to its Voronoi cell. Therefore, the shape of each cell is irregular, which is well suited for small cell networks. We assume universal frequency reuse, and thus there will be severe intercell interference. Due to limited backhaul capacity in small cell networks, joint transmission from multiple BSs \cite{Tanbourgi14,Nigam14} is not considered.

In this paper, we assume each BS serves at most one user at each time slot, i.e., intra-cell time division multiple access (TDMA) is adopted\footnote{Although single-user transmission is not necessarily the best option for multi-antenna transmission, our focus is on the interference nulling strategy and the derivation can be extended to other orthogonal multiple access methods, such as SDMA \cite{Li13_GC}.}. Due to the random locations of BSs and users, the number of users in each cell is random. For cells with no users, the BSs are called \emph{inactive BSs}, and they will not transmit any signal. Otherwise, the BS will be called an \emph{active BS} and will randomly choose one user in its cell to serve at each time slot. The probability that the typical BS is active is denoted as $p_{\rm a}$. Equivalently, $p_{\rm a}$ can be regarded as the ratio of the number of active BSs to the total number of BSs for each realization of $\Psi_b$ and $\Psi_u$. It has been shown that $p_{\rm a}$, as a function of the BS-user density ratio $\rho\triangleq\frac{\lambda_{b}}{\lambda_{u}}$, is given by \cite{Li14,Lee12}
\begin{equation} \label{eq:pa}
    p_{\rm a} \approx 1-\left(1+\frac{1}{c_0\rho}\right)^{-c_0},
\end{equation}
where $c_0=3.5$ is a constant related to the cell size distribution obtained through data fitting.

\subsection{User-Centric Intercell Interference Nulling}

\begin{figure}
    \begin{center}
    \scalebox{0.6}{\includegraphics{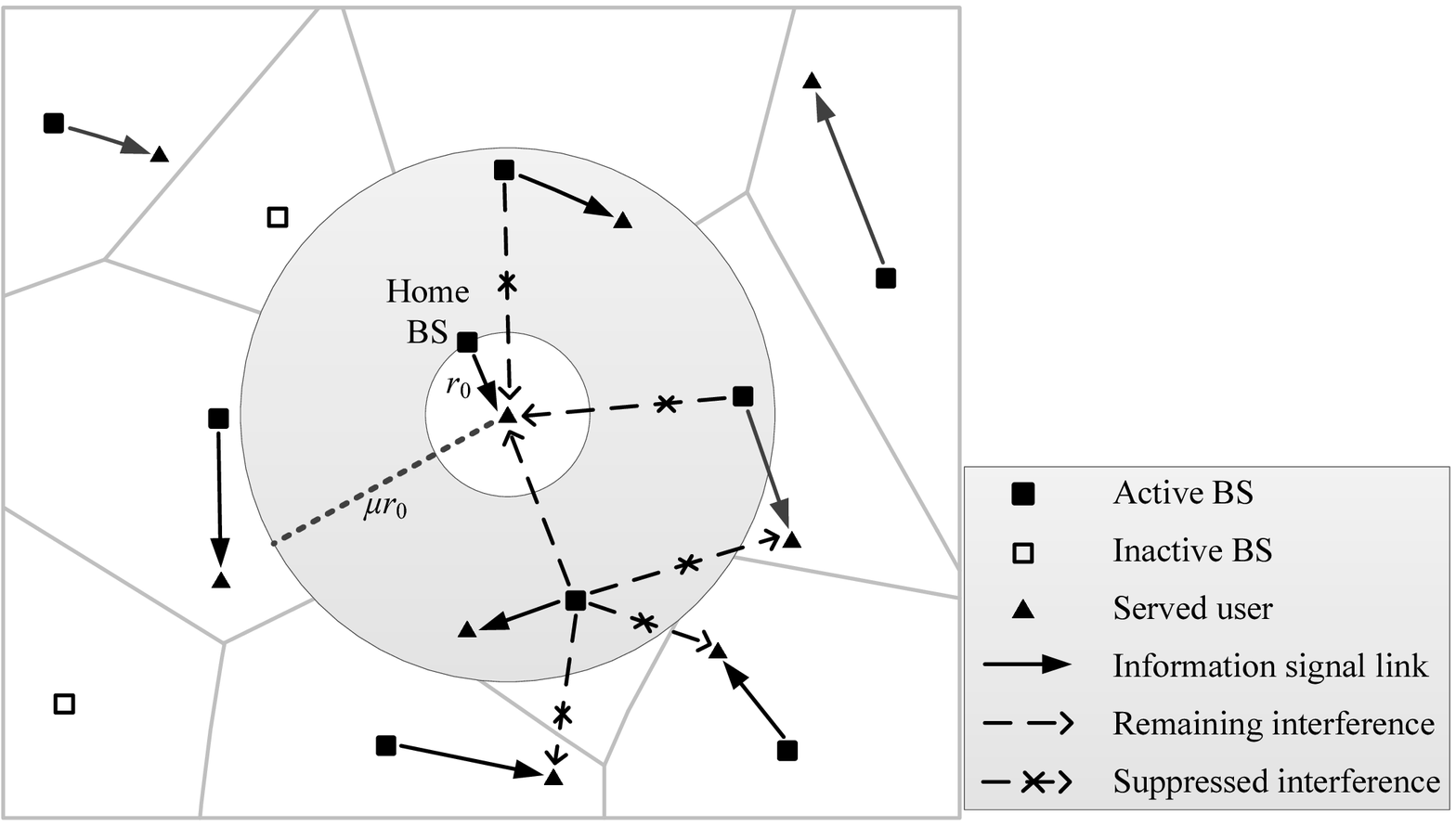}}
    \end{center}
    \caption{A sample network where BSs and users are distributed as two independent PPPs. The typical user is located at the origin, and the interfering BSs in the gray region will receive coordination requests from the typical user, but some of them may not be able to suppress interference due to the degrees of freedom constraint.}
    \label{fig:SystemModel}
\end{figure}

In the following, we will propose a user-centric intercell interference nulling strategy to suppress intercell interference for small cell networks. We propose that each served user will request a subset of interfering BSs for interference nulling. An interfering BS will be in this subset if the ratio of the average power received from this interfering BS to the average power received from the home BS is larger than a certain threshold, i.e., its interference is strong relative to the user's information signal power. Since each BS uses the same transmit power, the coordinating BSs can be determined by the relative distances to the interfering BSs and the home BS. Specifically, considering the typical user with distance $r_0$ to its home BS, it will request all the interfering BSs within distance $\mu r_0$ (where $\mu \geq 1$) for interference nulling. In the following, we will call $\mu r_0$ the \emph{IN range}, and the parameter $\mu$ the \emph{IN range coefficient}. Note that as the distance information is relatively easy to obtain, the proposed method to determine the IN range for each user incurs much less overhead than the ones based on instantaneous channel information. Moreover, as only the dominant interfering sources will be suppressed, it will lead to a more efficient utilization of the available radio resources, and better performance will be achieved.

Note that once determined, the value of $\mu$ is the same for all the users, i.e., the proposed strategy has a single design parameter. However, due to the random locations of BSs and users, the signal transmission distance $r_0$ is different in different cells, which means the area of interference coordination regions will be different for different users. Fig. \ref{fig:SystemModel} illustrates the BSs who will receive requests from the typical user, and all of them are within the annulus (the gray area) from radius $r_0$ to $\mu r_0$. Thus, the number of coordination requests received by a BS is a random variable, i.e., a BS may belong to multiple annuluses centered around different users. We denote the number of requests received by the BS located at coordinate $x$ as $K_x$, and denote the number of antennas at each BS as $M$. As $K_x$ is random and unbounded, it is possible that $K_x \geq M$. Due to the limited spatial degrees of freedom, each BS can handle at most $M-1$ requests \cite{Zhang09}. If a BS receives $K_x \geq M$ requests, we assume it will randomly choose $M-1$ users to suppress interference\footnote{Note that more sophisticated schemes to handle excess requests can be developed to further improve performance, but from the results shown in Section \ref{Sec:NumResults}, the improvement would be marginal since the value of $K_x$ is typically small for most BSs.}. This implies that it is possible for the requesting user to receive interference from the BSs within the annulus (as shown in Fig. \ref{fig:SystemModel}).

\begin{remark}[The effect of the IN range coefficient]
    Tuning the value of $\mu$ has conflicting effects: Increasing $\mu$ can suppress more nearby intercell interference. But the BSs will have less degrees of freedom for their own signal links, which will reduce the received information signal power. As a special case, $\mu=1$ implies a \emph{non-coordination scenario}, i.e., no interference nulling is employed in the network, and each active BS will serve its own user by single user beamforming. Our objective is to analytically evaluate the performance of the proposed coordination strategy and find the optimal $\mu$ to achieve the best performance.
\end{remark}

\subsection{Channel Model and Precoding Vectors}

We consider the typical user located at the origin $o$, served by its home BS at location $x_0$. This user will receive interference from the BSs outside the annulus and probably also from the BSs within the annulus. Let $\Psi_{b}^{\left(1\right)}$ denote the set of interfering BSs farther than $\mu r_0$, where $r_0=\left\Vert x_0\right\Vert$ and $\left\Vert \cdot\right\Vert $ is the vector norm. Let $\Psi_{b}^{\left(2\right)}$ denote the set of BSs who receive the request from the typical user but are unable to mitigate interference for this user. We assume Rayleigh fading channels, and denote the small-scale fading from the BS at location $x$ as $\mathbf{h}_{x} \in \mathcal{CN} \left(\mathbf{0}_{M\times1},\mathbf{I}_{M}\right)$. The large-scale path loss is modeled as $\left\Vert x\right\Vert^{-\frac{\alpha}{2}}$, where $\alpha>2$ represents the path loss exponent. Then, the received signal of this user is given by
\begin{equation} \label{eq:SignalModel}
    y_{o} = P_{t}^{\frac{1}{2}}r_{0}^{-\frac{\alpha}{2}}\mathbf{h}_{x_{0}}^{\ast}\mathbf{w}_{x_{0}}s_{x_{0}} +\sum_{x\in\Psi_{b}^{\left(1\right)}\cup\Psi_{b}^{\left(2\right)}}P_{t}^{\frac{1}{2}}\left\Vert x\right\Vert ^{-\frac{\alpha}{2}}\mathbf{h}_{x}^{\ast}\mathbf{w}_{x}s_{x}+n_{0},
\end{equation}
where $s_{x}\sim\mathcal{CN}\left(0,1\right)$ denotes the information symbol from the BS at $x$, $n_0\sim\left(0,\sigma_N^2\right)$ is the additive white Gaussian noise (AWGN), and $\mathbf{w}_x$ is the $M \times 1$ precoding vector for the BS at $x$.

In this paper we will adopt linear beamforming for interference nulling \cite{Zhang10,Jindal11}. We will first assume that the home BS and the interfering BSs that need to suppress interference to the user have perfect CSI, while the effect of imperfect CSI will be investigated in Section \ref{Sec:LimitedFeedback}.  We assume the typical user's home BS receives $K_{x_0}$ requests, and thus this BS will help $\min\left(K_{x_0},M-1\right)$ users to suppress interference. Denoting the channels of those requested users as $\mathbf{f}_{1},\ldots,\mathbf{f}_{\min\left(K_{x_0},M-1\right)}$, then the precoding vector $\mathbf{w}_{x_0}$ is given by
\begin{equation} \label{eq:w0}
    \mathbf{w}_{x_0}=\frac{\left(\mathbf{I}_M- \mathbf{F}\left(\mathbf{F}^{\ast}\mathbf{F}\right)^{-1}\mathbf{F}^{\ast}\right) {\mathbf{h}_{x_0}}} {\left\Vert \left(\mathbf{I}_M-\mathbf{F}\left(\mathbf{F}^{\ast}\mathbf{F}\right)^{-1}\mathbf{F}^{\ast}\right) {\mathbf{h}_{x_0}}\right\Vert },
\end{equation}
where $\mathbf{I}_M$ is the $M \!\times\! M$ identity matrix, and $\mathbf{F}=\left[\mathbf{f}_{1},\ldots,\mathbf{f}_{\min\left(K_{x_0},M-1\right)}\right]$.

From \eqref{eq:SignalModel} and \eqref{eq:w0}, the receive SINR of the typical user is given by
\begin{equation} \label{eq:SINR_ori}
    {\rm SINR}= \frac{P_{t}g_{0}r_{0}^{-\alpha}} {\sum_{x\in\Psi_{b}^{\left(1\right)}\cup\Psi_{b}^{\left(2\right)}} P_{t}g_{x}\left\Vert x\right\Vert ^{-\alpha}+\sigma_{N}^{2}},
\end{equation}
where $g_{0}\triangleq\left|\mathbf{h}_{x_0}^{\ast}\mathbf{w}_{x_0}\right|^{2}$ is the information signal channel gain, $g_{x}\triangleq\left|\mathbf{h}_{x}^{\ast}\mathbf{w}_{x}\right|^{2}$ is the interfering channel gain from the BS at $x$, and $\sigma_{N}^{2}$ is the noise power. It is shown in \cite{Jindal11} that in the perfect CSI case, $g_{0}$ is gamma distributed with shape parameter $M-\min\left(K_{x_0},M-1\right)$, i.e., $g_{0}\sim{\rm Gamma}\left[\max\left(M-K_{x_0},1\right),1\right]$, and the interfering channel gain $g_{x}$ is exponential with mean 1.

\subsection{Performance Metric and Key Approximations}

In this paper, we use the successful transmission probability to the typical user as the network performance metric, which is defined as $p_{\rm s}\triangleq \mathbb{P}\left({\rm SINR}\geq\hat\gamma\right)$, where $\hat\gamma$ is the SINR threshold. However, for the typical user, the distribution of receive SINR depends on $K_{x_0}$, i.e., the number of coordination requests received by its home BS. Thus, we denote the successful transmission probability to the user whose home BS receives $k$ requests as
\begin{equation}
    p_{{\rm s}}\left(k\right)=\mathbb{P}\left({\rm SINR}\geq\hat{\gamma}\mid K_{x_0}=k\right).
\end{equation}
Therefore, the average performance of the typical user is given by
\begin{equation} \label{eq:ps_sum}
    p_{\rm s}=\mathbb{E}_{K_{x_0}}\left[p_{\rm s}\left(K_{x_0}\right)\right]=\sum_{k=0}^{\infty}p_{\rm s}\left(k\right)p_K\left(k\right),
\end{equation}
where $p_K\left(k\right)$ is the probability mass function of $K_{x_0}$.

The reason of adopting $p_{\rm s}$ as the performance metric is that it can directly measure the average link reliability in the network. Moreover, the improvement of the successful transmission probability also reflects the improvement of the network throughput. Note that there are different types of throughput metrics focusing on different transmission schemes. For example, for the fixed rate transmission, the spatial throughput is given by $\lambda_bp_{\rm a}p_{\rm s}\log\left(1+\hat\gamma\right)$. On the other hand, if the BS can react quickly to the SINR condition and adjust its rate of transmission, then another type of spatial throughput, called the Shannon throughput, is defined as $\lambda_bp_{\rm a}\mathbb{E}\left[\log\left(1+\rm{SINR}\right)\right]$ \cite{Haenggi12}. For given BS and user densities, increasing the successful transmission probability $p_{\rm s}$ improves the throughput, no matter which metric is used. Thus, we will focus on the successful transmission probability in this paper.

As the performance analysis of the studied network is quite challenging, in the following we make a few key approximations. Firstly, since we consider small cell networks, which are interference-limited, we ignore the additive noise in the theoretical analysis. Secondly, we assume the numbers of users in different cells are independent (the same approximation has been used in \cite{Li14,Lee12}), and the numbers of requests received by different BSs are independent (i.e., $\left\{ K_{x}:\: x\in\Psi_{b}\right\} $ are independent random variables). These approximations simplify the analysis since the independent thinning of a PPP can be applied. Specifically, under such approximations, the set of active BSs is an independent thinning of $\Psi_b$. Thus, the density of $\Psi_b^{\left(1\right)}$ is $\lambda_1\left(x\right)=p_{{\rm a}}\lambda_{b}\mathbf{1}\left(\left\Vert x\right\Vert >\mu r_{0}\right)$, where $\mathbf{1}\left(\left\Vert x\right\Vert >\mu r_{0}\right)$ is the indicator function that equals 1 if $\left\Vert x\right\Vert >\mu r_{0}$ and 0 otherwise. Let $\varepsilon$ denote the probability that the BS has received the request from the user but is unable to null interference for this user. The density of $\Psi_b^{\left(2\right)}$ is then $\lambda_{2}\left(x\right)=\varepsilon p_{{\rm a}}\lambda_{b}\mathbf{1}\left(\left\Vert x\right\Vert \in\left[r_{0},\mu r_{0}\right]\right)$. To obtain an analytical expression of $\varepsilon$, consider an interfering BS from the annulus $\left[r_{0},\mu r_{0}\right]$ chosen uniformly at random. Besides the request from the typical user, assume it receives $\hat{K}$ more requests from other users. If $\hat{K}=\hat{k}\geq M-1$, then with probability $\frac{\hat{k}+1-\left(M-1\right)}{\hat{k}+1}$ this BS will not perform interference nulling for the typical user, as in this case this BS will randomly pick $M-1$ from $\hat{k}+1$ requests for interference nulling. To make the analysis tractable, we assume that the request from the typical user to this BS is independent of other users' situations, so that the probability mass function of $\hat{K}$ is approximated as $p_{K}\left(k\right)$. It follows that $\varepsilon$ can be approximated as
\begin{equation}
    \varepsilon \approx \sum_{k=M-1}^{\infty}\frac{k+1-\left(M-1\right)}{k+1}p_{K}\left(k\right).
\end{equation}

With the above approximations, the receive SINR of the typical user in \eqref{eq:SINR_ori} is simplified as
\begin{equation} \label{eq:SINR_def}
    {\rm SINR}=\frac{g_{0}r_{0}^{-\alpha}}{\sum_{x\in\Psi_{b}^{\left(1\right)}\cup\Psi_{b}^{\left(2\right)}} g_{x}\left\Vert x\right\Vert ^{-\alpha}}.
\end{equation}
In Section \ref{Sec:ps}, we will use \eqref{eq:SINR_def} to analyze the successful transmission probability. The accuracy of the approximations will be tested via simulations. For convenience, the key notations and symbols used in the paper are listed in Table \ref{tab:Notation}.

\begin{table}
    \caption{\label{tab:Notation}Key notations and symbols used in the paper}
    \begin{tabular}{|c|l|c|l|}
    \hline
    Symbol                   & Definition/Explanation                                     & Symbol                    & Definition/Explanation                                    \\ \hline
    $\lambda_b$, $\lambda_u$ & BS density, user density                                   & $\Psi_b^{\left(1\right)}$ & Set of interfering BSs farther than IN range              \\
    $\rho$                    & BS-user density ratio, i.e., $\frac{\lambda_b}{\lambda_u}$ & $\Psi_b^{\left(2\right)}$ & Set of BSs who receive the request from the typical user, \\
    $M$                      & \# of antennas in a BS                                 & ~                         & but are unable to mitigate interference to this user      \\
    $\alpha$                 & Path loss exponent ($\delta\triangleq2/\alpha$)            & $\Psi_b^{\left(3\right)}$ & Set of BSs who mitigate interference to the typical user  \\
    $\hat{\gamma}$           & SINR threshold                                             & $\varepsilon$             & Probability that a BS receives the request from a user,   \\
    $r_0$                    & Distance to the home BS                                    & ~                         & but is unable to mitigate interference for this user      \\
    $\mu$                    & IN range coefficient                                       & $p_{\rm a}$               & BS activity probability, determined by $\rho$             \\
    $K_x$                    & \# of requests received by the BS at $x$               & $p_K\left(k\right)$       & The probability mass function of $K_x$                    \\
    $B$                      & \# of feedback bits for one channel               & $p_{\rm s}\left(k\right)$ & The successful transmission probability to the user,           \\
    ~                        & vector                                                          & ~                         & whose home BS receives $k$ requests                       \\ \hline
    \end{tabular}
\end{table}

\section{Analysis of Successful Transmission Probability - The Perfect CSI Case} \label{Sec:ps}

It is shown from \eqref{eq:ps_sum} that the successful transmission probability is composed of $p_{\rm s}\left(k\right)$ and the distribution of $K_x$, i.e., $p_K\left(k\right)$. In this section, we will first derive $p_{\rm s}\left(k\right)$ and $p_K\left(k\right)$, which will then give an approximate expression of the successful transmission probability.

\subsection{The Expression of $p_{\rm s}\left(k\right)$}

In this subsection, we focus on the successful transmission probability to the user whose home BS receives $k$ requests. From the SINR expression in \eqref{eq:SINR_def}, $p_{\rm s}\left(k\right)$ is given by
\begin{equation}
    p_{{\rm s}}\left(k\right)=\mathbb{P}\left(\frac{g_{0}r_{0}^{-\alpha}} {\sum_{x\in\Psi_{b}^{\left(1\right)}\cup\Psi_{b}^{\left(2\right)}}g_{x}\left\Vert x\right\Vert ^{-\alpha}}\geq\hat{\gamma}\right).
\end{equation}
Since $g_0\sim{\rm Gamma}\left[\max\left(M-k,1\right),1\right]$,  using the cumulative distribution function of $g_0$, $p_{\rm s}\left(k\right)$ can be written as
\begin{equation}
    p_{{\rm s}}\left(k\right)=\mathbb{P}\left(g_{0}\geq \mathfrak{s}I\right)=\mathbb{E}_{\mathfrak{s}}\left[\sum_{n=0}^{\max\left(M-k,1\right)-1} \mathbb{E}_{I} \left[\frac{\left(\mathfrak{s}I\right)^{n}}{n!}e^{-\mathfrak{s}I}\right]\right],
\end{equation}
where $\mathfrak{s}\triangleq \hat{\gamma}r_0^\alpha$ and $I\triangleq\sum_{x\in\Psi_{b}^{\left(1\right)}\cup\Psi_{b}^{\left(2\right)}}g_{x}\left\Vert x\right\Vert ^{-\alpha}$. Note that for a fixed $\mathfrak{s}$, $\mathbb{E}_I\left[e^{-\mathfrak{s}I}\right]$ is the Laplace transform of $I$, denoted as $\mathcal{L}_I\left(\mathfrak{s}\right)$. Following the property of the Laplace transform, we have $\mathbb{E}_{I}\left[I^{n}e^{-\mathfrak{s}I}\right]= \left(-1\right)^{n}\mathcal{L}_{I}^{\left(n\right)}\left(\mathfrak{s}\right)$, where $\mathcal{L}_{I}^{\left(n\right)}\left(\mathfrak{s}\right)$ is the $n$th derivative of $\mathcal{L}_{I}\left(\mathfrak{s}\right)$. Then, we get
\begin{equation} \label{eq:ps_k_Laplace}
    p_{{\rm s}}\left(k\right) = \mathbb{E}_{\mathfrak{s}} \left[\sum_{n=0}^{\max\left(M-k,1\right)-1} \frac{\left(-\mathfrak{s}\right)^{n}}{n!} \mathcal{L}_{I}^{\left(n\right)}\left(\mathfrak{s}\right)\right].
\end{equation}

The major difficulty in the following derivation is to simplify the $n$th derivative of $\mathcal{L}_{I}\left(\mathfrak{s}\right)$, which is a common issue when dealing with multi-antenna transmission in the PPP network model \cite{Elsawy13}. In \cite{Li14}, a novel method was proposed to obtain a simple expression of the successful transmission probability. We follow a similar approach and  derive the successful transmission probability $p_{\rm s}\left(k\right)$, presented in the following proposition.

\begin{proposition} \label{Lemma:PsK}
The successful transmission probability to the user whose home BS receives $k$ requests is given by
\begin{equation} \label{eq:PsK}
    p_{{\rm s}}\left(k\right) = \left\Vert \left[\mathbf{I}_l+p_{{\rm a}}\mathbf{Q}_{l}\right]^{-1}\right\Vert _{1},
\end{equation}
where $\left\Vert\cdot\right\Vert_{1}$  is the $L_1$  matrix norm (i.e., $\left\Vert \mathbf{A}\right\Vert_{1}=\max_{1\leq j \leq n}\sum_{i=1}^{m}\left|a_{ij}\right|$ for $\mathbf{A}\in\mathbb{R}^{m\times n}$), $l=\max\left(M-k,1\right)$, $\mathbf{I}_l$ is the $l\times l$ identity matrix, and $\mathbf{Q}_{l}$ is a lower triangular Toeplitz matrix given by
\begin{equation}
    \mathbf{Q}_{l}=\left[\begin{array}{ccccc}
    q_{0}\\
    -q_{1} & q_{0}\\
    -q_{2} & -q_{1} & q_{0}\\
    \vdots &  & \ddots & \ddots\\
    -q_{l-1} & -q_{l-2} & \cdots & -q_{1} & q_{0}
    \end{array}\right]. \nonumber
\end{equation}
The elements of $\mathbf{Q}_{l}$ are given by
\begin{equation} \label{eq:q0}
    q_{0}=\hat{\gamma}^{\delta}\int_{\mu^{2}\hat{\gamma}^{-\delta}}^{\infty}\frac{du}{1+u^{1/\delta}} +\varepsilon\hat{\gamma}^{\delta} \int_{\hat{\gamma}^{-\delta}}^{\mu^{2}\hat{\gamma}^{-\delta}}\frac{du}{1+u^{1/\delta}},
\end{equation}
where $\delta\triangleq2/\alpha$, and for $i\geq1$,
\begin{equation} \label{eq:qi}
    q_{i}=\hat{\gamma}^{\delta} \int_{\mu^{2}\hat{\gamma}^{-\delta}}^{\infty} \frac{du}{\left(1+u^{-1/\delta}\right)\left(1+u^{1/\delta}\right)^{i}} +\varepsilon\hat{\gamma}^{\delta}\int_{\hat{\gamma}^{-\delta}}^{\mu^{2}\hat{\gamma}^{-\delta}} \frac{du}{\left(1+u^{-1/\delta}\right)\left(1+u^{1/\delta}\right)^{i}}.
\end{equation}
\end{proposition}
\begin{IEEEproof}
    See Appendix \ref{App:ps_k}.
\end{IEEEproof}

\subsection{The Approximate Expression of $p_{\rm s}$}

Based on the expression of $p_{\rm s}\left(k\right)$ in Proposition \ref{Lemma:PsK}, the successful transmission probability to the typical user can be obtained by substituting \eqref{eq:PsK} into \eqref{eq:ps_sum}. Hence,
\begin{equation} \label{eq:ps}
    p_{\rm s}=\sum_{k=0}^{M-1}\left\Vert \left[\mathbf{I}+p_{{\rm a}}\mathbf{Q}_{M-k}\right]^{-1} \right\Vert _{1}p_K\left(k\right) + \frac{1}{1+p_{\rm a}q_0} \sum_{k=M}^{\infty} p_K\left(k\right).
\end{equation}
However, the probability mass function of $K_x$ is difficult to obtain. In this subsection, we approximate the distribution of $K_x$ as Poisson by matching the mean. Note that even if the distribution of $K_x$ can be derived exactly, probably in a complicated form, the exact successful transmission probability is still difficult to obtain. Therefore, we resort to seeking a simple but tight approximation, which helps us to obtain a tractable expression of the successful transmission probability. Then, we can numerically obtain the optimal IN range coefficient $\mu$.

\begin{remark}
    Moment matching is a common method to obtain a tractable expression for complicated distributions since the exact results are usually difficult to obtain in the PPP network model. For example, in \cite{Singh13}, the authors used the first moment matching to approximate the area of a cell in heterogeneous cellular networks. In \cite{Tanbourgi14,Akoum13,Heath13,Haenggi09}, the Gamma distribution was used for approximating the residual interference by second order moment matching. It is a practical approach, and the results are tight in general.
\end{remark}

To approximate the distribution of $K_x$, we first obtain the first moment of $K_x$, i.e., $\bar{K}$, which is provided in the following lemma.

\begin{lemma} \label{Lemma:EK}
    The expected number of requests received by a BS is $\bar{K}=p_{\rm a}\left(\mu^2-1\right)$.
\end{lemma}
\begin{IEEEproof}
    See Appendix \ref{App:EK}.
\end{IEEEproof}

\begin{figure}
    \begin{center}
    \scalebox{0.8}{\includegraphics{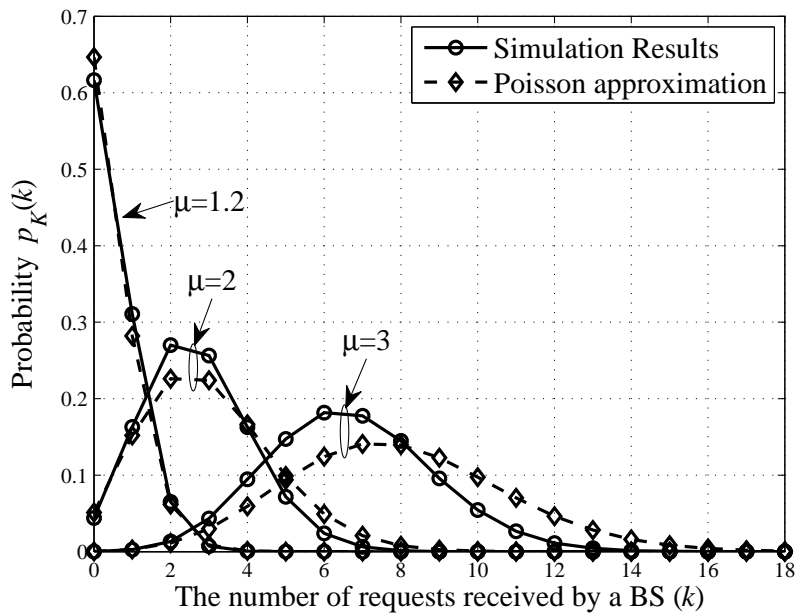}}
    \end{center}
    \caption{The probability mass function of $K_x$, i.e., $p_K\left(k\right)$, with $\rho=0.1$. }
    \label{fig:PDF_K}
\end{figure}

Based on Lemma \ref{Lemma:EK}, the probability mass function of $K_x$ is then approximated using the Poisson distribution as
\begin{equation} \label{eq:Appr_PK}
    p_K\left(k\right)=\mathbb{P}\left(K_x=k\right)\approx\frac{\left(\bar{K}\right)^k}{k!}e^{-\bar{K}}.
\end{equation}
Note that from \eqref{eq:Appr_PK}, the distribution of $K_x$ only depends on the BS-user density ratio $\rho$ and the IN range coefficient $\mu$. In Fig. \ref{fig:PDF_K}, we demonstrate the accuracy of the approximation, and it is shown that the approximation is more accurate for small values of $\mu$. In Section \ref{Sec:NumResults}, we will further test the impact of this approximation.

By substituting \eqref{eq:Appr_PK} into \eqref{eq:ps}, the approximate expression of the successful transmission probability to the typical user is given by
\begin{equation} \label{eq:ps_Approximation}
    p_{{\rm s}}\approx\sum_{k=0}^{M-1}\left\Vert \left[\mathbf{I}+p_{{\rm a}}\mathbf{Q}_{M-k}\right]^{-1}\right\Vert _{1}\frac{\left[p_{{\rm a}}\left(\mu^{2}-1\right)\right]^{k}}{k!}e^{-p_{{\rm a}}\left(\mu^{2}-1\right)}+\frac{\gamma\left[M,p_{{\rm a}}\left(\mu^{2}-1\right)\right]}{\left(M-1\right)!\left(1+p_{{\rm a}}q_{0}\right)},
\end{equation}
where $\gamma\left(a,b\right)$ is the lower incomplete Gamma function.

\begin{remark}[The effect of the BS and user densities]
It is apparent from \eqref{eq:ps_Approximation} that the effect of $\lambda_b$ and $\lambda_u$ on $p_{\rm s}$ is determined by the BS-user density ratio $\rho$.  In the following of this paper, we will change $\rho$ to investigate the effect of the BS density or the user density. Increasing $\rho$ can be viewed as increasing the BS density for a given user density, or equivalently, as decreasing the user density with a certain BS density.
\end{remark}

\begin{remark}[The non-coordination strategy]
Note that when $\mu=1$, \eqref{eq:ps_Approximation} becomes the exact expression, i.e., $p_{{\rm s}}=\left\Vert \left[\mathbf{I}_{M}+p_{{\rm a}}\mathbf{Q}_{M}\right]^{-1}\right\Vert _{1}$, where $q_0$ and $q_i$ are given in \eqref{eq:q0} and \eqref{eq:qi} with $\mu=1$ and $\varepsilon=0$. The result for this special case was obtained in \cite{Li14}. In the rest of the paper, we refer the performance of the non-coordination strategy as $p_{\rm s}$ in \eqref{eq:ps_Approximation} for $\mu=1$.
\end{remark}

\subsection{Performance Evaluation}

\begin{figure}
    \begin{center}
    \scalebox{0.8}{\includegraphics{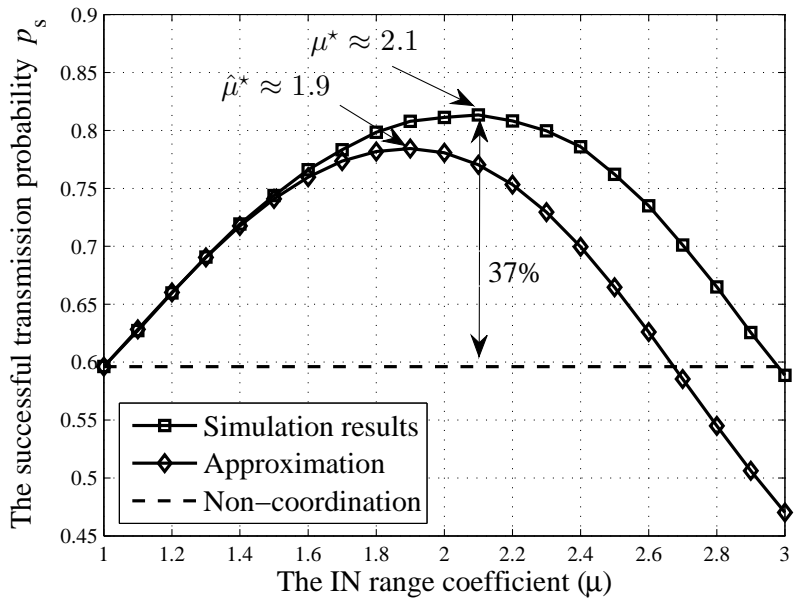}}
    \end{center}
    \caption{The successful transmission probability as a function of $\mu$, with $\lambda_b=10^{-3}{\rm m}^{-2}$, $\lambda_u=10^{-2}{\rm m}^{-2}$, $M=8$, $\alpha=4$ and $\hat{\gamma}=10$. The maximum performance gain of 37\% is the relative improvement from 60\% to 82\%.}
    \label{fig:ps_PCSIT}
\end{figure}

By now, we have obtained an approximation of the successful transmission probability using the user-centric intercell interference nulling strategy with a fixed $\mu$. We can then search for the optimal $\mu$ numerically, which is the $\mu$ that maximizes the successful transmission probability, i.e.,
\begin{equation}
    \mu^\star=\arg\max_{\mu}p_{{\rm s}}.
\end{equation}
In the rest of the paper, we will use $\mu^\star$ and $\hat{\mu}^\star$ to denote the optimal values obtained through simulation and based on the approximation in \eqref{eq:ps_Approximation}, respectively. A main benefit of our analytical approach is that $\hat{\mu}^\star$ can be found much more efficiently than $\mu^\star$, which requires extensive simulations. Next, we would like to examine the effectiveness of the proposed strategy and the tightness of the approximation.

In Fig. \ref{fig:ps_PCSIT}, we compare the simulation results with the approximation results, where the BS density is $\lambda_b=0.001$ per ${\rm m}^2$, user density is $\lambda_u=0.01$ per ${\rm m}^2$ and the SINR threshold is $\hat\gamma=10$. From Fig. \ref{fig:ps_PCSIT}, we can infer that selecting a proper IN range coefficient $\mu$ can greatly improve the network performance and that there exists an optimal $\mu$ to achieve the maximum successful transmission probability. Particularly, compared with the non-coordination scenario (i.e., $\mu=1$), using the user-centric intercell interference nulling with the optimal $\mu$ can improve the relative performance by about 37\%, which indicates the effectiveness of the proposed method. Moreover, by comparing the simulation results with the approximation, we find that the approximation result is lower than the simulation, and the approximation error increases with $\mu$. This is because the approximated $p_K\left( k\right)$ is less accurate when $\mu$ is large. However, it is also shown that the optimal IN range coefficient $\mu$ obtained from the approximation ($\hat{\mu}^\star\approx1.9$) is close to the optimal value from simulation ($\mu^\star\approx2.1$). As the curve of $p_{\rm s}$ is quite flat near $\mu^\star$, a small deviation of $\hat{\mu}^\star$ will only slightly affect $p_{\rm s}$, and thus we can obtain a near-optimal $\mu$ via the approximate expression. More results will be shown in Section \ref{Sec:NumResults} to confirm the tightness of the approximation.

\section{The Successful Transmission Probability with Limited Feedback} \label{Sec:LimitedFeedback}

The results in Section \ref{Sec:ps} were derived assuming perfect CSI. However, there will always be inaccuracy in the available CSI, which will degrade the performance. In this section, we consider the case where the active BSs will obtain quantized CSI through limited feedback, which is a common technique to provide CSI at the transmitter side \cite{Love08}.

\subsection{Limited Feedback Model}

With limited feedback, the channel direction information (CDI) is fed back using a quantization codebook known at both the transmitter and receiver \cite{Love08}. The quantization is chosen from a codebook of unit norm vectors of size $2^B$, where $B$ is the number of feedback bits for each channel. We assume that each user uses a different codebook to avoid getting the same quantization vector for different channels. The codebook for the typical user is denoted as $\mathcal{C}_{o}=\left\{ \bar{\mathbf{c}}_{j}:j=1,2,\ldots,2^{B}\right\} $, where the codewords are generated using random vector quantization (RVQ), i.e., each quantization vector $\bar{\mathbf{c}}_{j}$ is independently chosen from the isotropic distribution on the $M$ dimensional unit sphere \cite{Jindal06,Zhang10}. It has been shown in \cite{Jindal06,Akoum13} that RVQ can facilitate the analysis and provide performance close to the optimal quantization. Each user quantizes its CDI to the closest codeword, measured by the inner product. Therefore, the quantized CDI is
\begin{equation}
    \hat{\mathbf{h}}_{x} = \arg\max_{\bar{\mathbf{c}}_{j}\in\mathcal{C}_o} \left|\bar{\mathbf{h}}_{x}^{\ast}\bar{\mathbf{c}}_{j}\right|,
\end{equation}
where $\bar{\mathbf{h}}_{x}\triangleq\frac{\mathbf{h}_{x}}{\left\Vert \mathbf{h}_{x}\right\Vert }$ is the actual CDI. Then, the index of the quantized CDI $\hat{\mathbf{h}}_{x}$ is fed back with $B$ bits. In this paper, we assume the feedback channel is error-free and without delay. Thus, each active BS will use the quantized CDI of both the signal and interference channels to design its transmission vector.

For user-centric intercell interference nulling, each user not only needs to feed back CDI to its home BS, but also to the coordinating BSs. We assume that each user feeds back all the quantized CDI to its home BS, and then the home BS forwards the associated CDI to the corresponding BSs through backhaul connection. With the imperfect CSI at each BS, the received signal for the typical user (at the origin $o$) is given by
\begin{equation}
    y_{o}= P_{t}^{\frac{1}{2}}r_{0}^{-\frac{\alpha}{2}}\mathbf{h}_{x_{0}}^{\ast}\hat{\mathbf{w}}_{x_{0}}s_{x_{0}} +\sum_{x\in\Psi_{b}^{\left(1\right)}\cup\Psi_{b}^{\left(2\right)}\cup\Psi_{b}^{\left(3\right)}} P_{t}^{\frac{1}{2}}\left\Vert x\right\Vert ^{-\frac{\alpha}{2}}\mathbf{h}_{x}^{\ast}\hat{\mathbf{w}}_{x}s_{x}+n_{0},
\end{equation}
where  $\Psi_{b}^{\left(3\right)}$ denotes the set of interfering BSs who help this user to suppress interference. It can be shown that $\Psi_{b}^{\left(3\right)}$ is a non-homogeneous PPP with density $\ensuremath{\lambda_{3}\left(x\right)=\left(1-\varepsilon\right)p_{{\rm a}}\lambda_{b}\mathbf{1}\left(\left\Vert x\right\Vert \in\left[r_{0},\mu r_{0}\right]\right)}$. Note that in the perfect CSI case, the BSs in $\Psi_{b}^{\left(3\right)}$ do not cause interference to the user. However, with limited feedback, there is residual interference from these BSs due to the quantization error.  Moreover, the precoding vector of the BS at $x$, denoted as $\hat{\mathbf{w}}_x$, has the same expression as \eqref{eq:w0} but is designed based on quantized CDI. Therefore, the receive SINR can be written as
\begin{equation} \label{eq:SINR_LF_def}
    {\rm SINR}= \frac{\hat{g}_{0}r_{0}^{-\alpha}} {\sum_{x\in\Psi_{b}^{\left(1\right)}\cup\Psi_{b}^{\left(2\right)}\cup\Psi_{b}^{\left(3\right)}} \hat{g}_{x}\left\Vert x\right\Vert ^{-\alpha}},
\end{equation}
where the equivalent channel gain is given as $\hat{g}_x=\left|\mathbf{h}_{x}^{\ast}\hat{\mathbf{w}}_{x}\right|^{2}$. Compared with the SINR expression \eqref{eq:SINR_def} in the perfect CSI case, it is clear that due to limited feedback, there is another part of interference, which is  $\sum_{x\in\Psi_{b}^{\left(3\right)}}\hat{g}_{x}\left\Vert x\right\Vert ^{-\alpha}$. Moreover, $\hat{g}_0$ no longer follows the gamma distribution with scale parameter 1, and its parameter will depend on the number of feedback bits $B$. In the next subsection, we will derive the distribution of the channel gains and then obtain the expression of the successful transmission probability based on \eqref{eq:SINR_LF_def}.

\subsection{Expression of $p_{\rm s}$ with Limited Feedback}

To determine the successful transmission probability, we first need to get the distribution of channel gains with limited feedback. Since the precoding vector $\hat{\mathbf{w}}_{x}$ is independent with the channel from the BS at $x\in\Psi_{b}^{\left(1\right)}\cup\Psi_{b}^{\left(2\right)}$, $\hat{g}_x$ is still exponential, i.e., $\hat{g}_x\sim{\rm Exp}\left(1\right)$ for $x\in\Psi_{b}^{\left(1\right)}\cup\Psi_{b}^{\left(2\right)}$. However, due to the quantization error, the distributions of the information channel gain $\hat{g}_0$ and the residual interference channel gain $\hat{g}_x$ for $x\in\Psi_{b}^{\left(3\right)}$ will change, which are given in the following lemma.

\begin{lemma} \label{Lemma:ChannelGains}
    Given the number of feedback bits for one channel vector as $B$, the distribution of the information channel gain $\hat{g}_0$ can be approximated as $\hat{g}_{0}\sim{\rm Gamma}\left[\max\left(M-k,1\right),\kappa_{0}\right]$, where $M$ is the number of BS antennas, $k$ is the number of requests received by this BS, and $\kappa_{0}\triangleq 1-2^{B}\beta\left(2^{B},\frac{M}{M-1}\right)$ where $\beta\left(x,y\right)=\frac{\Gamma\left(x\right)\Gamma\left(y\right)}{\Gamma\left(x+y\right)}$ is the Beta function.

    Moreover, the residual interference channel gain $\hat{g}_x$ for $x\in\Psi_{b}^{\left(3\right)}$ can be approximated as $\hat{g}_{x}\sim{\rm Exp}\left(1/{\kappa_{I}}\right)$, where $\kappa_{I}=2^{B}\beta\left(2^{B},\frac{M}{M-1}\right)=1-\kappa_0$ is the quantization distortion. 
\end{lemma}
\begin{IEEEproof}
    The proof is similar to Lemma 5 in \cite{Zhang10}.
\end{IEEEproof}

Based on the SINR expression \eqref{eq:SINR_LF_def}, the successful transmission probability is given by
\begin{equation} \label{eq:ps_k_LF_ori}
    p_{{\rm s,LF}}\left(k\right)= \mathbb{P}\left(\frac{\hat{g}_{0}r_{0}^{-\alpha}} {\sum_{x\in\Psi_{b}^{\left(1\right)}\cup\Psi_{b}^{\left(2\right)}\cup\Psi_{b}^{\left(3\right)}} \hat{g}_{x}\left\Vert x\right\Vert ^{-\alpha}}\ge\hat{\gamma}\right),
\end{equation}
where ``LF'' represents limited feedback. Then, following the same procedure of  Proposition \ref{Lemma:PsK} and using the distributions of $\hat{g}_{0}$ and $\hat{g}_{x}$ in Lemma \ref{Lemma:ChannelGains}, the successful transmission probability to the typical user is given in the following proposition.

\begin{proposition} \label{Lemma:PsK_LF}
    With limited feedback, the successful transmission probability to the user whose home BS receives $k$ requests is given by
    \begin{equation} \label{eq:ps_k_LF}
        p_{{\rm s,LF}}\left(k\right)=\left\Vert \left[\mathbf{I}_{l}+p_{{\rm a}}\mathbf{Q}_{l}\right]^{-1}\right\Vert _{1},
    \end{equation}
    where $l=\max\left(M-k,1\right)$ and $\mathbf{Q}_{l}$ has the same structure as in Proposition \ref{Lemma:PsK}, with $q_0$ and $q_i$ replaced by
    \begin{equation} \label{eq:q0_LF}
        q_{0,{\rm LF}}= \left(\!\frac{\hat{\gamma}}{\kappa_{0}}\!\right)^{\!\!\delta} \!\!\int_{\mu^{2}\left(\!\frac{\hat{\gamma}}{\kappa_{0}}\!\right)^{-\delta}}^{\infty} \frac{du}{1+u^{\frac{1}{\delta}}} +\varepsilon\left(\!\frac{\hat{\gamma}}{\kappa_{0}}\!\right)^{\!\!\delta} \!\!\int_{\left(\!\frac{\hat{\gamma}}{\kappa_{0}}\!\right)^{-\delta}}^{\mu^{2} \left(\!\frac{\hat{\gamma}}{\kappa_{0}}\!\right)^{-\delta}} \!\!\!\!\! \frac{du}{1+u^{\frac{1}{\delta}}} +\left(1-\varepsilon\right)\left(\!\frac{\kappa_{I}}{\kappa_{0}}\hat{\gamma}\!\right)^{\!\!\delta} \!\!\int_{\left(\!\frac{\kappa_{I}}{\kappa_{0}}\hat{\gamma}\!\right)^{-\delta}}^{\mu^{2} \left(\!\frac{\kappa_{I}}{\kappa_{0}}\hat{\gamma}\!\right)^{-\delta}} \!\!\!\!\! \frac{du}{1+u^{\frac{1}{\delta}}},
    \end{equation}
    \begin{eqnarray} \label{eq:qi_LF}
        q_{i,{\rm LF}} &=& \left(\frac{\hat{\gamma}}{\kappa_{0}}\right)^{\delta} \int_{\mu^{2}\left(\frac{\hat{\gamma}}{\kappa_{0}}\right)^{-\delta}}^{\infty} \frac{du}{\left(1+u^{-\frac{1}{\delta}}\right)\left(1+u^{\frac{1}{\delta}}\right)^{i}} +\varepsilon\left(\frac{\hat{\gamma}}{\kappa_{0}}\right)^{\delta} \int_{\left(\frac{\hat{\gamma}}{\kappa_{0}}\right)^{-\delta}}^{\mu^{2} \left(\frac{\hat{\gamma}}{\kappa_{0}}\right)^{-\delta}} \!\!\! \frac{du}{\left(1+u^{-\frac{1}{\delta}}\right)\left(1+u^{\frac{1}{\delta}}\right)^{i}} \nonumber\\
        &+& \left(1-\varepsilon\right) \left(\frac{\kappa_{I}}{\kappa_{0}}\hat{\gamma}\right)^{\delta} \int_{\left(\frac{\kappa_{I}}{\kappa_{0}}\hat{\gamma}\right)^{-\delta}}^{\mu^{2} \left(\frac{\kappa_{I}}{\kappa_{0}}\hat{\gamma}\right)^{-\delta}} \frac{du}{\left(1+u^{-\frac{1}{\delta}}\right)\left(1+u^{\frac{1}{\delta}}\right)^{i}}.
    \end{eqnarray}
\end{proposition}
\begin{IEEEproof}
    See Appendix \ref{App:ps_k_LF}.
\end{IEEEproof}
By substituting \eqref{eq:ps_k_LF} into \eqref{eq:ps_sum}, we can obtain the final expression of the successful transmission probability with limited feedback.

\begin{remark}
    By comparing the expressions of the successful transmission probability of the perfect CSI case and the limited feedback case, we can observe that the only terms changed are $q_0$ and $q_i$. Moreover, the quantization distortion $\kappa_I$ decreases when increasing the number of feedback bits $B$, and $\kappa_I \rightarrow0$ when $B\rightarrow \infty$. This means that $q_{0,{\rm LF}}$ and $q_{i,{\rm LF}}$ in \eqref{eq:q0_LF} and \eqref{eq:qi_LF} will converge to $q_0$ and $q_i$ of the perfect CSI case as $B$ increases.
\end{remark}

\subsection{Performance Evaluation}

\begin{figure}
    \begin{center}
    \scalebox{0.8}{\includegraphics{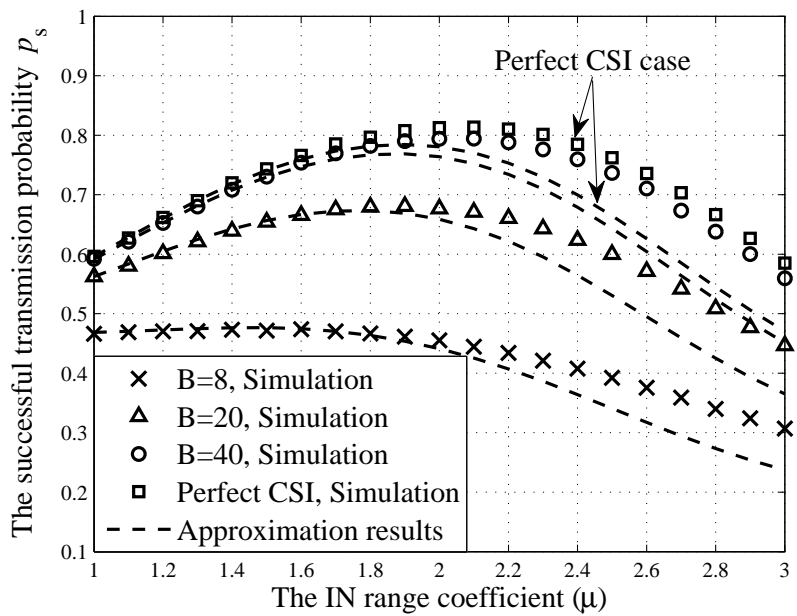}}
    \end{center}
    \caption{The successful transmission probability as a function of $\mu$, with $\lambda_b=10^{-3}{\rm m}^{-2}$, $\lambda_u=10^{-2}{\rm m}^{-2}$, $M=8$, $\alpha=4$ and $\hat{\gamma}=10$. The dashed lines are the approximation results for $B=8,20, 40$, and the perfect CSI case, respectively.}
    \label{fig:ps_LF}
\end{figure}

In Fig. \ref{fig:ps_LF}, we show the effect of the number of feedback bits $B$ on the successful transmission probability. We see that with limited feedback, the successful transmission probability is still a quasi-concave function with respect to the IN range coefficient $\mu$, i.e., $p_{\rm s}$ will first increase and then decrease when $\mu$ increases. Moreover, similar to the perfect CSI case, the approximation is more accurate when $\mu$ is small, but for different values of $B$, the optimal $\mu$ obtained via the approximation is close to the one via simulation. Thus, the approximate result can help to optimize the proposed interference nulling strategy.

\section{Numerical Results} \label{Sec:NumResults}

In this section, we will compare the proposed strategy with other interference nulling methods, and then present some numerical results to provide guidelines for practical system design.

\subsection{Performance Comparison of Different Interference Nulling Strategies}

First, we will compare the proposed interference nulling strategy with other interference nulling methods. One method for comparison is similar to that used in \cite{Xia13}, where each user will request a fixed-number ($N$) of BSs for interference nulling, denoted as \emph{number based ICIN}. Particularly, each user requests $N$ nearest interfering BSs to suppress interference. But if the BS receives more than $M-1$ requests, it will randomly choose $M-1$ users to mitigate interference. The other method for comparison is the random BS clustering method, proposed in \cite{Akoum13}.

\begin{figure}
    \begin{center}
    \scalebox{0.8}{\includegraphics{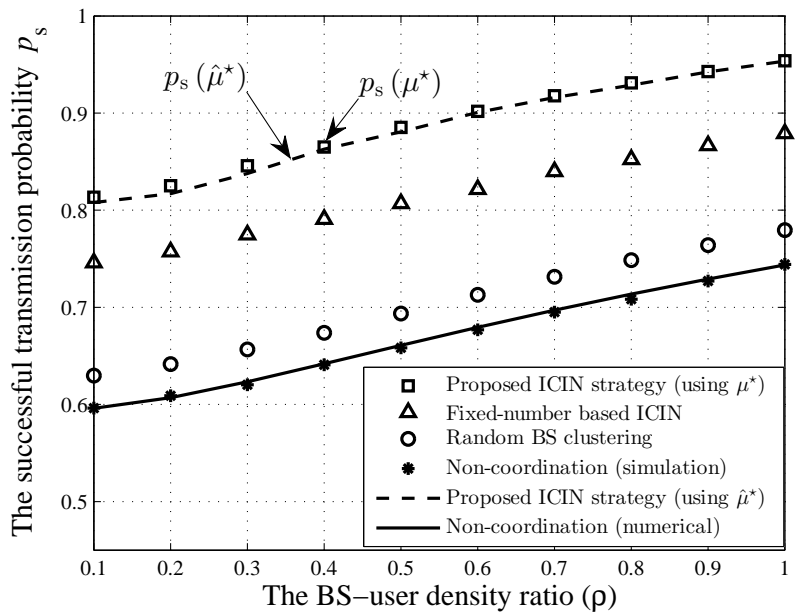}}
    \end{center}
    \caption{The successful transmission probability with different BS-user density ratios, with $\lambda_b=10^{-3}{\rm m}^{-2}$, $M=8$, $\alpha=4$ and $\hat{\gamma}=10$, where ``ICIN'' stands for intercell interference nulling. The dashed line is obtained through simulation using the approximated optimal $\mu$ (i.e., $\hat\mu^\star$) from \eqref{eq:ps_Approximation}, while the square is obtained by using the optimal $\mu$  (i.e., $\mu^\star$) in simulation.}
    \label{fig:ps_rho}
\end{figure}

Fig. \ref{fig:ps_rho} shows the comparison results for different BS-user density ratios. Note that for all methods, we use the optimal value of the key parameter, i.e., for the proposed strategy, we use the optimal IN range coefficient $\mu$. For the fixed-number based ICIN, we optimize $N$ to obtain the maximum $p_{\rm s}$. And for random BS clustering, we find the optimal cluster size\footnote{The optimal $N$ of the number based ICIN and the optimal cluster size for random BS clustering can only be obtained via simulation as no analytical expression of $p_s$ is available for these two cases.}. Moreover, the successful transmission probability without coordination is presented as the baseline. From Fig. \ref{fig:ps_rho}, we can find that: 1) User-centric coordination methods significantly outperform the BS clustering method, and the proposed method performs better than the fixed-number based ICIN.  2) Using the approximated optimal $\mu$ (denoted as $\hat{\mu}^\star$) provides performance very close to that using simulation to search the optimal $\mu$ (denoted as $\mu^\star$), so it can be used in practice. 3) As $\rho$ increases\footnote{Note that previous works such as \cite{Akoum13,Huang13,Tanbourgi14,Nigam14,Zhang13,Xia13} focused on the case $p_{\rm a}=1$, which could not capture the effect of the user distribution.}, the successful transmission probability increases, and it appears that the performance gaps between different methods stay constant.

The superior performance of the proposed coordinated strategy is because it can more effectively identify the dominant interference for each user, while the other two coordination methods are not adaptive to each user's interference situation. Furthermore, we find that the optimal $\mu$ in Fig. \ref{fig:ps_rho} is about $2$, from which we can derive the average number of requests $\bar{K}$ to be around $2\sim3$, so most BSs can well handle the requests with the available spatial degrees of freedom. This also confirms that our proposed method is practical since the IN range is not large, and thus the amount of signaling overhead will be acceptable.

\subsection{Guidelines for Practical Network Deployment}

\begin{figure}
    \begin{center}
    \scalebox{0.8}{\includegraphics{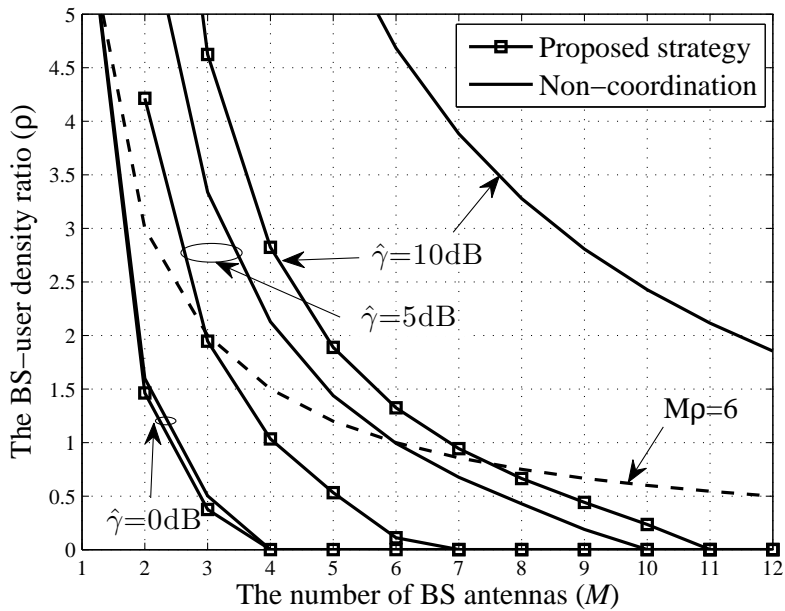}}
    \end{center}
    \caption{The minimal BS-user density ratio required to achieve $p_{\rm s}=0.9$, with different numbers of BS antennas, with $\alpha=4$. For the proposed strategy, the approximated optimal $\mu$ (i.e., $\hat{\mu}^\star$) is used to obtain $p_{\rm s}$, then we find the minimal $\rho$ to achieve $p_{\rm s}=0.9$. The dashed line is the reference line, on which all the points have the same value of $M\rho$.}
    \label{fig:Tradeoff}
\end{figure}

Next, we shall provide some design guidelines for the practical network deployment with our proposed interference nulling strategy. We will consider two different options to improve the network performance, i.e., to deploy more BSs or to increase the number of BS antennas. The effects of these two approaches are shown in Fig. \ref{fig:Tradeoff}. In this figure, for a given value of $M$, we obtain the minimal BS-user density ratio $\rho$ required to achieve the successful transmission probability of 0.9 with the SINR threshold as 0 dB, 5 dB and 10 dB, respectively. The approximated optimal $\mu$ (i.e., $\hat{\mu}^\star$) is used for the proposed interference nulling strategy. The following interesting and insightful observations can be made: 1) When the SINR threshold $\hat{\gamma}$ is small, the performance of the proposed strategy is similar with the performance of the non-coordination strategy. It can be found that the optimal $\mu$ tends to 1 when $\hat{\gamma}$ decreases. 2) When $\hat{\gamma}$ is large, the advantage of the proposed interference nulling strategy is significant. For example, with $M=6$ and $\hat{\gamma}=10$ dB, it can achieve the same performance as the non-coordination strategy with only $1/3$ of the BS density; while with the BS-user density ratio as 2, it achieves the same performance at $M=5$ instead of $M=12$. It means that the deployment cost can be greatly reduced with the proposed interference nulling strategy to meet the requirement of high data rate transmission.  3) The number of BS antennas plays a more important role than the BS density. If we fix the total number of antennas per unit area, e.g., fix $\rho M=6$ as in Fig. \ref{fig:Tradeoff}, it is shown that increasing $M$ can improve the supported SINR threshold, which implies that co-located BS antennas can support higher data rate requirement\footnote{This conclusion depends on the actual transmission strategy, and a full comparison between co-located and distributed antenna deployment is left to future work.}.

\subsection{The Impact of Imperfect CSI}

\begin{figure}
    \begin{center}
    \scalebox{0.8}{\includegraphics{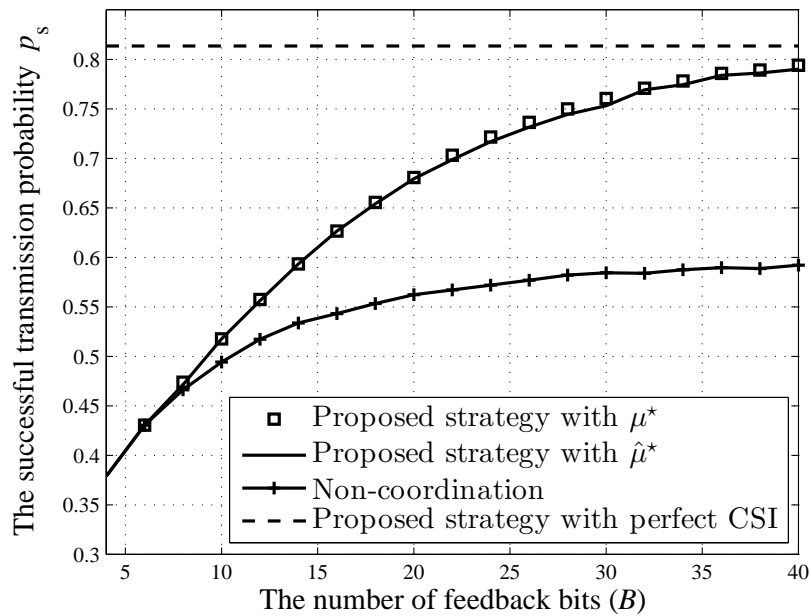}}
    \end{center}
    \caption{The successful transmission probability with different $B$, with $\lambda_b=10^{-3}{\rm m}^{-2}$, $\lambda_u=10^{-2}{\rm m}^{-2}$, $M=8$, $\alpha=4$ and $\hat{\gamma}=10$. }
    \label{fig:ps_B}
\end{figure}

So far we have demonstrated the effectiveness of the proposed user-centric intercell interference nulling strategy, especially for the high SINR requirement (i.e., $\hat\gamma$ is high). Next, we will investigate the effect of the limited feedback on the performance.

In Fig. \ref{fig:ps_B}, we evaluate the effect of the number of feedback bits $B$, where the performances of cooperative and non-cooperative systems are compared with different values of $B$. We find that when the number of feedback bits $B$ increases, the successful transmission probability will approach the perfect CSI case. However, if $B$ is not sufficiently large, using interference nulling has a similar performance with the non-coordination strategy. This is because when $B$ is small, the quantization error is large, which will limit the performance of interference nulling. Thus, a sufficient number of bits are required to quantize each channel vector in order to exploit the performance gains of interference nulling, e.g., $B\geq10$ for the network considered in Fig. \ref{fig:ps_B}.

\section{Conclusions} \label{Sec:Conclusions}

In this paper, we proposed a novel interference nulling strategy for downlink small cell networks, which we refer to as user-centric intercell interference nulling. By comparing the interference power with the received information signal power, the dominant interfering BSs  for each user can be identified effectively, which brings performance gains compared with other coordination methods, such as random BS clustering.  Specifically, it was demonstrated that satisfactory user performance can be achieved in small cell networks, with intercell interference suppressed by effective coordination strategies, supported by a sufficient number of antennas at each BS and with accurate CSI. Moreover, random spatial network models were proved to be a powerful tool to analyze and design cooperative cellular networks, in which it is critical to consider the spatial distributions of both BSs and users.

With a low implementation complexity and a higher performance gain, the proposed interference coordination strategy will have wide applications in cellular networks. The user-centric approach can be easily applied to other interference management methods that are performed in the time or frequency domain. One limitation of the current work is that we only considered the same type of BSs. The extension to more general heterogeneous cellular networks (HetNets) therefore requires further investigation. Moreover, the consideration of other transmission techniques, such as multiuser MIMO, will be an interesting research direction.

\appendix

\subsection{Proof of Proposition \ref{Lemma:PsK}} \label{App:ps_k}

To derive $p_{\rm s}\left(k\right)$ based on \eqref{eq:ps_k_Laplace}, we start from the Laplace transform of $I$ conditioning on a fixed $\mathfrak{s}$, given by
\begin{equation}
    \mathcal{L}_{I}\left(\mathfrak{s}\right)=\mathbb{E} \left[\exp\left(-\mathfrak{s}\sum_{x\in\Psi_{b}^{\left(1\right)}\cup\Psi_{b}^{\left(2\right)}} g_{x}\left\Vert x\right\Vert ^{-\alpha}\right)\middle| \mathfrak{s}\right].
\end{equation}
Since the channel gains $g_x$ are independent and identically distributed (i.i.d.) exponential random variables over $x$, the above equality can be written as
\begin{equation} \label{eq:Laplace_Div_forLF}
    \mathcal{L}_{I}\left(\mathfrak{s}\right)= \mathbb{E}\left[\prod_{x\in\Psi_{b}^{\left(1\right)}\cup\Psi_{b}^{\left(2\right)}} \frac{1}{1+\mathfrak{s}\left\Vert x\right\Vert ^{-\alpha}}\middle| \mathfrak{s} \right].
\end{equation}
Then, the Laplace transform of $I$ can be derived using the probability generating functional (PGFL) \cite{Haenggi12}, which is given as
\begin{equation} \label{eq:LaplaceFinal}
    \mathcal{L}_{I}\left(\mathfrak{s}\right) = \exp\left\{ -\pi\lambda_{b}p_{{\rm a}} \left[\int_{\mu^{2}r_{0}^{2}}^{\infty}\!\left(\!1-\frac{1}{1+\mathfrak{s} u^{-\frac{\alpha}{2}}} \!\right)du +\varepsilon\!\int_{r_{0}^{2}}^{\mu^{2}r_{0}^{2}} \!\left(\!1-\frac{1}{1+\mathfrak{s} u^{-\frac{\alpha}{2}}}\!\right)du\right]\right\}.
\end{equation}
Based on \eqref{eq:LaplaceFinal}, the $n$th derivative of $\mathcal{L}_{I}\left(\mathfrak{s}\right)$ with respect to $\mathfrak{s}$ can be written according to the following recursive form
\begin{eqnarray} \label{eq:Laplace_Div}
    \mathcal{L}_{I}^{\left(n\right)}\!\!\left(\mathfrak{s}\right) = \pi\lambda_{b}p_{{\rm a}} \!\!\sum_{i=0}^{n-1} \! \left(\!\!\!\!\begin{array}{c}
    n\!-\!1\\
    i
    \end{array}\!\!\!\!\right)\!\! \mathcal{L}_{I}^{\left(i\right)}\!\!\left(\mathfrak{s}\right) \left(n-i\right)! \! \left[\!\int_{\mu^{2}r_{0}^{2}}^{\infty} \!\! \frac{\left(-u^{-\frac{\alpha}{2}}\right)^{n-i}du} {\left(1\!+\!\mathfrak{s}u^{-\frac{\alpha}{2}}\right)^{n-i+1}} + \varepsilon\!\! \int_{r_{0}^{2}}^{\mu^{2}r_{0}^{2}} \!\!\! \frac{\left(-u^{-\frac{\alpha}{2}}\right)^{n-i}du} {\left(1\!+\!\mathfrak{s}u^{-\frac{\alpha}{2}}\right)^{n-i+1}}\right].
\end{eqnarray}

Denote $a_n=\frac{\left(-\mathfrak{s}\right)^n}{n!}\mathcal{L}_{I}^{\left(n\right)}\left(\mathfrak{s}\right)$ and substitute $\mathfrak{s}=\hat{\gamma}r_0^\alpha$ into \eqref{eq:LaplaceFinal}, then we have
\begin{equation}
    a_0=\mathcal{L}_{I}\left(\mathfrak{s}\right)=\exp\left(-\pi\lambda_b p_{\rm a}r_0^2 q_0 \right),
\end{equation}
where $q_0$ is given in \eqref{eq:q0}. Similarly, by substituting $\mathfrak{s}=\hat{\gamma}r_0^\alpha$ into \eqref{eq:Laplace_Div}, we get for $n\geq1$,
\begin{equation}
  a_n=\pi\lambda_b p_{\rm a}r_0^2 \sum_{i=0}^{n-1}\frac{n-i}{n}q_{n-i}a_i,
\end{equation}
where $q_i$ is presented in \eqref{eq:qi}. Note that $q_0$ and $q_i$ can be expressed as the Gauss hypergeometric functions \cite{Li14}. Since $a_0$ and $a_i$ have the same recursive structure as in \cite{Li14}, using the results derived in \cite{Li14}, the successful transmission probability can be obtained, as shown in \eqref{eq:PsK}.

\subsection{Proof of Lemma \ref{Lemma:EK}} \label{App:EK}

Denote the number of requests sent from the served user located at $y$ as $N_y$, which is a random variable due to the random distance between users and BSs. One basic equality is that the total number of requests sent by all served users should be equal to the total number of requests received by all active BSs. Since one active BS will only serve one user at each time slot, the expectation of $N_y$ is equal with the expectation of $K_x$, i.e., $\bar{N}=\bar{K}$. Therefore, to obtain $\bar{K}$, we can instead derive the distribution of $N_y$ and obtain $\bar{N}$.

Assuming the distance from the typical user to its home BS is $r_0$, then the number of interfering BSs ($N_o$) in the annulus from radius $r_0$ to $\mu r_0$  is Poisson distributed because interfering BSs follow a PPP with density $p_{\rm a}\lambda_b$. Denote the area of the annulus as $\mathcal{A}$, which is $\mathcal{A}=\pi\left(\mu r_{0}\right)^{2}-\pi r_{0}^{2}$, then the distribution of $N_o$ for a fixed $r_0$ is given by
\begin{equation} \label{eq:APP_PN_con_r0}
    \mathbb{P}\left(N_o=n\mid r_{0}\right)=\frac{\left[\lambda_b p_{{\rm a}}\mathcal{A}\right]^{n}}{n!}e^{-\lambda_b p_{{\rm a}}\mathcal{A}}.
\end{equation}
On the other hand, the transmission distance $r_0$ has the following distribution \cite{Haenggi05},
\begin{equation} \label{eq:APP_r0}
    f_{r_0}\left(r\right)=2\pi\lambda_b r e^{-\pi\lambda_b r^{2}}.
\end{equation}
Therefore, the unconditional distribution of $N_o$ is written as
\begin{equation} \label{eq:APP_PN_un_r0}
    \mathbb{P}\left(N_o=n\right) = \int_0^\infty \mathbb{P}\left(N_o=n\mid r_{0}\right)f\left(r_{0}\right)dr_0
    = \left[1+\frac{1}{p_{{\rm a}}\left(\mu^{2}-1\right)}\right]^{-n}\frac{1}{1+p_{{\rm a}}\left(\mu^{2}-1\right)}.
\end{equation}
Furthermore, the expectation of $N_o$ can be derived as $\bar{N}=p_{{\rm a}}\left(\mu^{2}-1\right)$, which also gives $\bar{K}$.

\subsection{Proof of Proposition \ref{Lemma:PsK_LF}} \label{App:ps_k_LF}

The successful transmission probability in \eqref{eq:ps_k_LF_ori} can be expressed as
\begin{eqnarray}
  p_{{\rm s,LF}}\left(k\right) &=& \mathbb{P}\left(\hat{g}_{0}\ge\hat{\gamma}r_{0}^{\alpha}I_{{\rm LF}}\right) = \mathbb{E}_{\hat{\mathfrak{s}}}\left[\sum_{n=0}^{\max\left(M-k,1\right)-1}\mathbb{E}_{I_{{\rm LF}}}\left[\frac{\left(\hat{\mathfrak{s}}I_{{\rm LF}}\right)^{n}}{n!}e^{-\hat{\mathfrak{s}}I_{{\rm LF}}}\right]\right] \nonumber\\
  &=& \mathbb{E}_{\hat{\mathfrak{s}}} \left[\sum_{n=0}^{\max\left(M-k,1\right)-1}\frac{\left(-\hat{\mathfrak{s}}\right)^{n}}{n!} \mathcal{L}_{I_{{\rm LF}}}^{\left(n\right)}\left(\hat{\mathfrak{s}}\right)\right],
\end{eqnarray}
where $\hat{\mathfrak{s}}\triangleq\frac{\hat{\gamma}r_{0}^{\alpha}}{\kappa_{0}}$, $I_{{\rm LF}} \triangleq \sum_{x\in\Psi_{b}^{\left(1\right)} \cup\Psi_{b}^{\left(2\right)} \cup\Psi_{b}^{\left(3\right)}} \hat{g}_{x}\left\Vert x\right\Vert ^{-\alpha}$, and $\mathcal{L}_{I_{{\rm LF}}}^{\left(n\right)}\left(\hat{\mathfrak{s}}\right)$ denotes the $n$th derivative of Laplace transform of $I_{\rm LF}$ with respect to a fixed $\hat{\mathfrak{s}}$. Similar to \eqref{eq:Laplace_Div_forLF}, $\mathcal{L}_{I_{{\rm LF}}}\left(\hat{\mathfrak{s}}\right)$ is given by
\begin{equation}
    \mathcal{L}_{I_{{\rm LF}}}\left(\hat{\mathfrak{s}}\right)= \mathbb{E}\left[\prod_{x\in\Psi_{b}^{\left(1\right)} \cup\Psi_{b}^{\left(2\right)}} \frac{1}{1+\hat{\mathfrak{s}}\left\Vert x\right\Vert ^{-\alpha}} \prod_{x\in\Psi_{b}^{\left(3\right)}} \frac{1}{1+\kappa_{I}\hat{\mathfrak{s}} \left\Vert x\right\Vert ^{-\alpha}} \middle|\hat{\mathfrak{s}}\right].
\end{equation}
Using the PGFL \cite{Haenggi12}, the Laplace transform of $I_{\rm LF}$ is then written as
\begin{equation}
    \mathcal{L}_{I_{{\rm LF}}}\left(\hat{\mathfrak{s}}\right)= \exp\left\{ -\pi\lambda_{b}p_{{\rm a}}\left[\int_{\mu^{2}r_{0}^{2}}^{\infty} \frac{\hat{\mathfrak{s}}u^{-\frac{\alpha}{2}}du}{1+\hat{\mathfrak{s}}u^{-\frac{\alpha}{2}}} + \varepsilon\!\int_{r_{0}^{2}}^{\mu^{2}r_{0}^{2}} \!\! \frac{\hat{\mathfrak{s}}u^{-\frac{\alpha}{2}}du}{1+\hat{\mathfrak{s}}u^{-\frac{\alpha}{2}}} + \!\left(1-\varepsilon\right)\!\int_{r_{0}^{2}}^{\mu^{2}r_{0}^{2}} \!\! \frac{\kappa_{I}\hat{\mathfrak{s}}u^{-\frac{\alpha}{2}}du} {1+\kappa_{I}\hat{\mathfrak{s}}u^{-\frac{\alpha}{2}}}\right]\right\}.
\end{equation}
Then, following the same procedure as in Appendix \ref{App:ps_k}, we can obtain the successful transmission probability shown in \eqref{eq:ps_k_LF}. 

\bibliographystyle{IEEEtran}
\bibliography{IEEEabrv,Ref}

\begin{thebibliography}{10}
\providecommand{\url}[1]{#1}
\def\UrlFont{\rmfamily}
\providecommand{\newblock}{\relax}
\providecommand{\bibinfo}[2]{#2}
\providecommand\BIBentrySTDinterwordspacing{\spaceskip=0pt\relax}
\providecommand\BIBentryALTinterwordstretchfactor{4}
\providecommand\BIBentryALTinterwordspacing{\spaceskip=\fontdimen2\font plus
\BIBentryALTinterwordstretchfactor\fontdimen3\font minus
  \fontdimen4\font\relax}
\providecommand\BIBforeignlanguage[2]{{%
\expandafter\ifx\csname l@#1\endcsname\relax
\typeout{** WARNING: IEEEtran.bst: No hyphenation pattern has been}%
\typeout{** loaded for the language `#1'. Using the pattern for}%
\typeout{** the default language instead.}%
\else
\language=\csname l@#1\endcsname
\fi
#2}}

\bibitem{Cisco14}
Cisco, ``Cisco visual network index: Global mobile traffic forecast update,
  2013-2018,'' pp. 1--40, Feb. 2014.

\bibitem{Hwang13}
I.~Hwang, B.~Song, and S.~Soliman, ``A holistic view on hyper-dense
  heterogeneous and small cell networks,'' \emph{IEEE Commun. Mag.}, vol.~51,
  no.~6, pp. 20--27, Jun. 2013.

\bibitem{Andrews11}
J.~G. Andrews, F.~Baccelli, and R.~K. Ganti, ``A tractable approach to coverage
  and rate in cellular networks,'' \emph{IEEE Trans. Commun.}, vol.~59, no.~11,
  pp. 3122--3134, Nov. 2011.

\bibitem{Li14}
C.~Li, J.~Zhang, and K.~B. Letaief, ``Throughput and energy efficiency analysis
  of small cell networks with multi-antenna base stations,'' \emph{IEEE Trans.
  Wireless Commun.}, vol.~13, no.~5, pp. 2505--2517, May 2014.

\bibitem{Gesbert10}
D.~Gesbert, S.~Hanly, H.~Huang, S.~Shamai~Shitz, O.~Simeone, and W.~Yu,
  ``Multi-cell {MIMO} cooperative networks: A new look at interference,''
  \emph{IEEE J. Sel. Areas Commun.}, vol.~28, no.~9, pp. 1380--1408, Sept.
  2010.

\bibitem{Zhang09}
J.~Zhang, R.~Chen, J.~G. Andrews, A.~Ghosh, and R.~W. Heath, ``Networked {MIMO}
  with clustered linear precoding,'' \emph{IEEE Trans. Wireless Commun.},
  vol.~8, no.~4, pp. 1910--1921, Apr. 2009.

\bibitem{LopezPerez11}
D.~L\'{o}pez-P\'{e}rez, I.~G\"{u}ven\c{c}, G.~de~la Roche, M.~Kountouris,
  T.~Quek, and J.~Zhang, ``Enhanced intercell interference coordination
  challenges in heterogeneous networks,'' \emph{IEEE Wireless Commun.},
  vol.~18, no.~3, pp. 22--30, Jun. 2011.

\bibitem{Foschini06}
G.~Foschini, K.~Karakayali, and R.~Valenzuela, ``Coordinating multiple antenna
  cellular networks to achieve enormous spectral efficiency,'' \emph{IEEE Proc.
  Commun.}, vol. 153, no.~4, pp. 548--555, Aug. 2006.

\bibitem{Shamai04}
S.~Shamai, O.~Somekh, and B.~M. Zaidel, ``Multi-cell communications: An
  information theoretic perspective,'' in \emph{Proc. Joint Workshop on Commun.
  and Coding (JWCC)}, Florence, Italy, Oct. 2004.

\bibitem{Huh12}
H.~Huh, A.~Tulino, and G.~Caire, ``Network {MIMO} with linear zero-forcing
  beamforming: Large system analysis, impact of channel estimation, and
  reduced-complexity scheduling,'' \emph{IEEE Trans. Inf. Theory}, vol.~58,
  no.~5, pp. 2911--2934, May 2012.

\bibitem{Bjornson13}
E.~Bj{\"o}rnson and E.~Jorswieck, ``Optimal resource allocation in coordinated
  multi-cell systems,'' \emph{Found. and Trends in Commun. and Inf. Theory},
  vol.~9, no. 2-3, pp. 113--381, Jan. 2013.

\bibitem{Zhang10}
J.~Zhang and J.~G. Andrews, ``Adaptive spatial intercell interference
  cancellation in multicell wireless networks,'' \emph{IEEE J. Sel. Areas
  Commun.}, vol.~28, no.~9, pp. 1455--1468, Dec. 2010.

\bibitem{Bhagavatula11}
R.~Bhagavatula and R.~Heath, ``Adaptive bit partitioning for multicell
  intercell interference nulling with delayed limited feedback,'' \emph{IEEE
  Trans. Signal Process.}, vol.~59, no.~8, pp. 3824--3836, Aug. 2011.

\bibitem{Zhang11}
J.~Zhang, J.~G. Andrews, and K.~B. Letaief, ``Optimizing training and feedback
  for spatial intercell interference cancellation,'' in \emph{Proc. IEEE Global
  Telecommun. Conf. (GLOBECOM)}, Houston, TX, Dec. 2011, pp. 1--5.

\bibitem{Simeone09}
O.~Simeone, O.~Somekh, H.~Poor, and S.~Shamai, ``Local base station cooperation
  via finite-capacity links for the uplink of linear cellular networks,''
  \emph{IEEE Trans. Inf. Theory}, vol.~55, no.~1, pp. 190--204, Jan. 2009.

\bibitem{Shamai01}
S.~Shamai and B.~Zaidel, ``Enhancing the cellular downlink capacity via
  co-processing at the transmitting end,'' in \emph{Proc. IEEE Veh. Technol.
  Conf. (VTC)}, vol.~3, Rhodes, Greece, May 2001, pp. 1745--1749.

\bibitem{Papadogiannis08}
A.~Papadogiannis, D.~Gesbert, and E.~Hardouin, ``A dynamic clustering approach
  in wireless networks with multi-cell cooperative processing,'' in \emph{Proc.
  IEEE Int. Conf. Commun. (ICC)}, Beijing, China, May 2008, pp. 4033--4037.

\bibitem{Xu11}
J.~Xu, J.~Zhang, and J.~Andrews, ``On the accuracy of the {Wyner} model in
  cellular networks,'' \emph{IEEE Trans. Wireless Commun.}, vol.~10, no.~9, pp.
  3098--3109, Sept. 2011.

\bibitem{Guo13}
A.~Guo and M.~Haenggi, ``Spatial stochastic models and metrics for the
  structure of base stations in cellular networks,'' \emph{IEEE Trans. Wireless
  Commun.}, vol.~12, no.~11, pp. 5800--5812, Nov. 2013.

\bibitem{Dhillon12}
H.~Dhillon, R.~Ganti, F.~Baccelli, and J.~Andrews, ``Modeling and analysis of
  {K}-tier downlink heterogeneous cellular networks,'' \emph{IEEE J. Sel. Areas
  Commun.}, vol.~30, no.~3, pp. 550--560, Apr. 2012.

\bibitem{Cheung12}
W.~C. Cheung, T.~Quek, and M.~Kountouris, ``Throughput optimization, spectrum
  allocation, and access control in two-tier femtocell networks,'' \emph{IEEE
  J. Sel. Areas Commun.}, vol.~30, no.~3, pp. 561--574, Apr. 2012.

\bibitem{Singh13}
S.~Singh, H.~S. Dhillon, and J.~G. Andrews, ``Offloading in heterogeneous
  networks: Modeling, analysis, and design insights,'' \emph{IEEE Trans.
  Wireless Commun.}, vol.~12, no.~5, pp. 2484--2497, May 2013.

\bibitem{Li13_GC}
C.~Li, J.~Zhang, and K.~B. Letaief, ``Performance analysis of {SDMA} in
  multicell wireless networks,'' in \emph{Proc. IEEE Global Telecommun. Conf.
  (GLOBECOM)}, Atlanta, GA, Dec. 2013.

\bibitem{Elsawy13}
H.~Elsawy, E.~Hossain, and M.~Haenggi, ``Stochastic geometry for modeling,
  analysis, and design of multi-tier and cognitive cellular wireless networks:
  A survey,'' \emph{IEEE Commun. Surveys \& Tutorials}, vol.~15, no.~3, pp.
  996--1019, 2013.

\bibitem{Wang14}
H.~Wang, X.~Zhou, and M.~Reed, ``Coverage and throughput analysis with a
  non-uniform small cell deployment,'' \emph{IEEE Trans. Wireless Commun.},
  2014, to appear.

\bibitem{Akoum13}
S.~Akoum and R.~W. Heath, ``Interference coordination: Random clustering and
  adaptive limited feedback,'' \emph{IEEE Trans. Signal Process.}, vol.~61,
  no.~7, pp. 1822--1834, Apr. 2013.

\bibitem{Huang13}
K.~Huang and J.~Andrews, ``An analytical framework for multicell cooperation
  via stochastic geometry and large deviations,'' \emph{IEEE Trans. Inf.
  Theory}, vol.~59, no.~4, pp. 2501--2516, Apr. 2013.

\bibitem{Tanbourgi14}
R.~Tanbourgi, S.~Singh, J.~G. Andrews, and F.~K. Jondral, ``A tractable model
  for non-coherent joint-transmission base station cooperation,'' \emph{{\rm
  submitted to} IEEE Trans. Wireless Commun.}, Jan. 2014, available:
  http://arxiv.org/abs/1308.0041v2.

\bibitem{Nigam14}
G.~Nigam, P.~Minero, and M.~Haenggi, ``Coordinated multipoint joint
  transmission in heterogeneous networks,'' \emph{{\rm submitted to} IEEE
  Trans. Commun.}, Feb. 2014, available:
  \url{http://www.nd.edu/~mhaenggi/pubs/tcom14.pdf}.

\bibitem{Zhang13}
X.~Zhang and M.~Haenggi, ``A stochastic geometry analysis of inter-cell
  interference coordination and intra-cell diversity,'' \emph{{\rm submitted
  to} IEEE Trans. Wireless Commun.}, 2013, available:
  http://arxiv.org/abs/1403.0012.

\bibitem{Xia13}
P.~Xia, C.-H. Liu, and J.~Andrews, ``Downlink coordinated multi-point with
  overhead modeling in heterogeneous cellular networks,'' \emph{IEEE Trans.
  Wireless Commun.}, vol.~12, no.~8, pp. 4025--4037, Aug. 2013.

\bibitem{Lee12}
S.~Lee and K.~Huang, ``Coverage and economy of cellular networks with many base
  stations,'' \emph{IEEE Commun. Lett.}, vol.~16, no.~7, pp. 1038--1040, Jul.
  2012.

\bibitem{Jindal11}
N.~Jindal, J.~G. Andrews, and S.~Weber, ``Multi-antenna communication in ad hoc
  networks: Achieving {MIMO} gains with {SIMO} transmission,'' \emph{IEEE
  Trans. Commun.}, vol.~59, no.~2, pp. 529--540, Feb. 2011.

\bibitem{Haenggi12}
M.~Haenggi, \emph{Stochastic geometry for wireless communications}.\hskip 1em
  plus 0.5em minus 0.4em\relax Cambridge University Press, 2012.

\bibitem{Heath13}
R.~W. Heath, M.~Kountouris, and T.~Bai, ``Modeling heterogeneous network
  interference using {Poisson} point processes,'' \emph{IEEE Trans. Signal
  Process.}, vol.~61, no.~16, pp. 4114--4126, Aug. 2013.

\bibitem{Haenggi09}
M.~Haenggi and R.~K. Ganti, \emph{Interference in large wireless
  networks}.\hskip 1em plus 0.5em minus 0.4em\relax Now Publishers Inc., 2009.

\bibitem{Love08}
D.~J. Love, R.~W. Heath, V.~K.~N. Lau, D.~Gesbert, B.~D. Rao, and M.~Andrews,
  ``An overview of limited feedback in wireless communication systems,''
  \emph{IEEE J. Sel. Areas Commun.}, vol.~26, no.~8, pp. 1341--1365, Oct. 2008.

\bibitem{Jindal06}
N.~Jindal, ``{MIMO} broadcast channels with finite-rate feedback,'' \emph{IEEE
  Trans. Inf. Theory}, vol.~52, no.~11, pp. 5045--5060, Nov. 2006.

\bibitem{Haenggi05}
M.~Haenggi, ``On distances in uniformly random networks,'' \emph{IEEE Trans.
  Inf. Theory}, vol.~51, no.~10, pp. 3584--3586, Oct. 2005.

\end{thebibliography}

\end{document}